\documentclass[twocolumn,prl,superscriptaddress,showpacs,letterpaper]{revtex4} 
\usepackage{graphicx} 
\usepackage{amsmath,amssymb}
\usepackage{dcolumn}

\begin{document} 
\title{Fluctuation and dissipation of work in a Joule experiment.} 
\author{B. Cleuren}
\affiliation{Hasselt University - B-3590 Diepenbeek, Belgium}
\author{C. Van den Broeck}
\affiliation{Hasselt University - B-3590 Diepenbeek, Belgium}
\author{R. Kawai}
\affiliation{Department of Physics, University of Alabama at Birmingham,
Birmingham, Alabama 35294, USA} 

\begin{abstract}
We elucidate the connection between various fluctuation theorems by a microcanonical version of the Crooks relation. We derive the microscopically exact expression for the work distribution in an idealized Joule experiment, namely for an object, convex but otherwise of arbitrary shape, moving at constant speed through an ideal gas. Analytic results are compared with molecular dynamics simulations of a hard disk gas.
\end{abstract}
\date{\today}

\maketitle
Microscopic time-reversibility implies, in a system at equilibrium, the basic
symmetry of  detailed balance,  stating that any process and its time reverse
occur equally frequently.  In the {\it linear} regime outside equilibrium this
property entails, as Onsager has first shown, a relation between fluctuation and
dissipation, since in this regime one\ cannot distinguish between the average
regression following an external perturbation or an equilibrium fluctuation.
Over the past decade, time-reversibility of deterministic or stochastic dynamics
has been shown to imply relations between fluctuation and dissipation  in
systems {\it far} from equilibrium, taking the form of a number of intriguing
equalities, the fluctuation theorem \cite{evans93,gallavotti95}, the Jarzynski
equality \cite{jarzynski97a} and the Crooks relation \cite{crooks99}. 
In this letter, we want to stress the relation between these results and discuss
their relevance by a microscopically exact study of a Joule experiment. 
\begin{figure}[tb]
\includegraphics[width=2.5in]{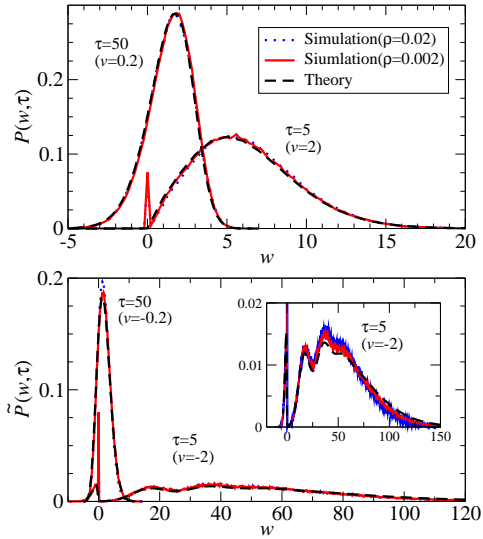}
\caption{Work distributions for a triangular object moving to the $+x$ 
(upper panel) and $-x$ direction (lower
panel). Inset: detail of the multiple peak structure for $\tau=5$.}
\label{fig:pw}
\end{figure}

Our theoretical starting point will be the derivation of a microcanonical version of
the Crooks relation. This result has the advantage that we can consider from the
onset an isolated system,  thereby dispensing with the need for considering a
heat bath. In the limit of an infinitely large system, we recover the three
above mentioned equalities. The validity and experimental observability of these
relations  and their interconnection can be discussed from the exact result for
distribution of work when moving a convex object through an ideal gas. 

Consider an isolated system at time $t=0$ in microcanonical equilibrium at
energy $E$. Its Hamiltonian depends on a control parameter, which is varied
during the  time interval $[0,t]$ following a specified protocol. An initial
position  $x_i$ of the system in phase space will evolve according to
Hamiltonian dynamics into a final position $x_f$. The  corresponding initial and
final values of the Hamiltonian are denoted by  $H_i(x_i)$  and $H_f(x_f)$,
respectively.  During this process, an amount of work $W=H_f(x_f)-H_i(x_i)$  is
delivered to the system.  Due to the microcanonical sampling of the initial
state from the energy shell $H_i=E$, the work $W$ is a random variable with the
following probability density:
\begin{align}
&P_{E}(W)=\langle \delta(H_f(x_f)-H_i(x_i)-W) \rangle_E \notag\\
&=\frac{\int dx_i \delta(H_i(x_i)-E)
\delta(W-H_f(x_f)+H_i(x_i))}{\Omega_i(E)},
\label{eq:PW+}
\end{align}
where $\Omega_i(E)=\int dx_i \delta(H_i(x_i)-E)=\exp\{S_i(E)/k_B\}$ is the
volume of the energy shell and $S_i(E)$ is the entropy at the initial
equilibrium state. Consider now the time-reversed protocol. A phase point
$\tilde{x}_f$, where the
tilde refers to velocity inversion, will evolve in time to the final phase space
position $\tilde{x}_i$.  We now average over a microcanonical sampling in the
energy shell $H_f(\tilde{x}_f)=E+W$. The probability distribution
$\tilde{P}_{E+W}(-W)$ for a work $-W$ in this time reversed protocol is
given by:
\begin{align}
&\tilde{P}_{E+W}(-W)=\langle
\delta(H_f(\tilde{x}_f)-H_i(\tilde{x}_i)-W) \rangle_{E+W} \notag\\
&=\frac{\int d\tilde{x}_f \delta(H_f(\tilde{x}_f)-E-W)
\delta(H_f(\tilde{x}_f)-H_i(\tilde{x}_i)-W)}{\Omega_f(E+W)}.
\label{eq:PW-}
\end{align}
Since the Jacobian for the transformation from $dx_i $ to $d\tilde{x}_f$ is one,
the integrals in Eqs. (\ref{eq:PW+}) and (\ref{eq:PW-}) are identical, and the
following 
microcanonical Crooks relation follows:
\begin{equation}\label{crooksmicro}
\frac{P_{E}(W)}{\tilde{P}_{E+W}(-W)}=\frac{\Omega_f(E+W)}{\Omega_i(E)}=e^{{\frac
{S_f(E+W)-S_i(E)}{k_B}}}.
\end{equation}
In an appropriate thermodynamic limit, entailing  $E \rightarrow \infty$, the
work distributions converge to functions $P(W)$ and $\tilde{P}(-W)$, independent
of the energy of the system, while  the temperature  $T$, $\partial S/\partial E
=1/T$, is a well defined constant (i.e.,  same for the initial and final
microcanonical distribution). Since $\Delta F=\Delta E-T \Delta S$ and $\Delta
E=W$ the internal energy difference, one recovers the canonical Crooks
relation\cite{crooks99}:
\begin{equation}
\frac{P(W)}{\tilde{P}(-W)}=\exp\left \{\frac{\Delta
S}{k_B}\right\}=e^{\beta(W-\Delta F)},
\label{eq:crooks}
\end{equation}
from which the Jarzynski relation $\langle \exp(-\beta W)\rangle = \exp(-\beta \Delta F)$ follows by integration. Note that in the aforementioned thermodynamic limit, entailing $W/E \rightarrow 0$,
$\Delta F=W-T (S_f(E+W)-S_i(E)) \rightarrow -T(S_f(E)-S_i(E))$ is the free energy difference between final and initial state at same energy $E$,  independent of $W$.
Note also that for a protocol of
asymptotically long duration $t\rightarrow \infty$, that leads the system into a
nonequilibrium steady state, one can write  $\Delta S=t \sigma$, where $\sigma$
is the entropy production per time unit,  while $W=\Delta F+T \Delta S \approx
Tt \sigma$ . Eq. (\ref{eq:crooks}) then can be rewritten  $P(\sigma)/\tilde{P}(-\sigma)\sim
\exp(t \sigma/k_B)$, which in the particular case of a time-symmetric schedule with  $P=\tilde{P}$ reduces to the Evans-Cohen-Gallavotti fluctuation theorem \cite{evans93,gallavotti95}.

\begin{figure}[b]
\includegraphics[width=2.5in]{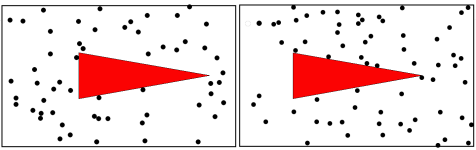}
\caption{Snapshots for a triangular object moving to the left (lhs panel) and to
right (rhs panel). Note the void behind the object when it moves in the
direction of its arrow (rhs panel). } \label{fig:snapshots}
\end{figure}

As an application of the above result, we turn to an exactly solvable microscopic model of a Joule experiment: an ideal gas at equilibrium in an infinitely large container receives an amount of mechanical energy $W$ by moving a closed convex body through it during a time duration $t$ at a fixed speed $V$ along  a fixed horizontal axis $x$ (see also Fig. \ref{fig:snapshots}).
We will calculate explicitly the probability distribution $P(W)$ for this work.  For
simplicity and for comparison with molecular dynamics, we restrict ourselves to
a two-dimensional system, with vertical axis $y$.  The shape of the object is
completely specified by the circumference $S$ and the form factor $F(\theta)$,
with  $F(\theta) d\theta$ defined as the fraction of the circumference with
polar angle between $\theta$ and $\theta + d\theta$, the angle being measured
counterclockwise from the $x$-axis (see Ref.\cite{meurs04}). When a gas particle
 hits the object, the amount of work $\Delta W$ supplied by the external force
is equal to the increase in kinetic energy of the particle. The post-collisional
speed is found in terms of the (precollisional) speed $\vec{v}=(v_{x},v_{y})$
from the conservation of total energy and total momentum in the $x$-direction.
The resulting work contribution is $\Delta W =-2mV\sin^{2}\theta(v_{x}-V-v_{y}\cot \theta)$. Note that this quantity is a random variable, through its dependence on the speed of the incoming particle and of the inclination $\theta$ of the impact point. In the case of an ideal gas at equilibrium (or in the limit of a extremely dilute gas, with the mean free path of the object much larger that its linear dimension, the so-called large Knudsen number regime), the subsequent collisions are independent random events. Hence the total work $W(t)$ after a time $t$, being the sum of uncorrelated identically distributed $\Delta W$'s, is a stochastic process with independent increments. The time evolution of the work distribution $P(W,t)$ is described by the following Master equation:
\begin{equation}
\partial_{t}P(W,t)=\int_{-\infty}^{+\infty} T(\Delta
W)(e^{-\Delta W \partial_{W}}-1)P(W,t) d\Delta W.
\label{me}
\end{equation}
The probability per unit time $T(\Delta W)$ for a change in $W$ by an amount
$\Delta W$ can be calculated following the basic methods of kinetic theory, and
is given by:
\begin{equation}
\begin{split}
&T(\Delta W)=\int_{0}^{2\pi} SF(\theta)d\theta
\int_{-\infty}^{+\infty}  \int_{-\infty}^{+\infty} d v_{x} d v_{y} \\ 
&\times \rho
\phi(v_{x},v_{y}) H[(\vec{V}-\vec{v}).\vec{e}_{\perp}(\theta)]
\vert (\vec{V}-\vec{v}).\vec{e}_{\perp} (\theta)\vert \\
&\times \delta[\Delta W + 2mV\sin^{2}\theta (v_{x}-V-v_{y}\cot
\theta)],
\end{split}
\end{equation}
where $H$ denotes the Heaviside function, $\vec{V}=(V,0)$,
$\vec{e}_{\perp}(\theta)$ is the vector orthogonal to the circumference at the
orientation $\theta$, while the delta functions picks out the incoming speeds
that give rise to the requested work contribution $\Delta W$. The gas is
characterized by the (uniform) density $\rho$ and the Maxwellian velocity
distribution $\phi(v_{x},v_{y})$ at temperature $T$. The solution of the Master
equation is found by Fourier transform, and is most easily expressed in terms of
the following dimensionless variables:
\begin{equation}
w=\beta W, \quad v=V\left(\frac{\beta m}{2}\right)^{1/2}, \quad
\tau=\frac{S\rho t}{(2 \beta m)^{1/2}},
\end{equation}
being the work and speed measured in terms of the thermal energy and speed of
the gas particles, and the time in terms of the average time between collisions.
The cumulant generating function $G(q)=\log \langle e^{-iqw}\rangle$ reads:
\begin{align}
G(q)&=\log \int_{-\infty}^{+\infty} e^{-iqw}P(w,\tau) dw
=\sum_{n=1}^{\infty}\frac{(-iq)^{n}}{n!}\kappa_{n} \notag\\
&=\tau \int_{0}^{2\pi} d\theta F(\theta)v\sin\theta
\biggl\{\Bigl(\text{erf}[(1-2iq)v\sin\theta] +1 \Bigr ) \biggr . \notag \\
& \biggl . \times (1-2iq) e^{-4q(i+q)v^{2}\sin^{2}\theta}-1-\text{erf}
(v\sin\theta) \biggr\}.
\label{gq}
\end{align}
Note that, as expected, the work distribution is invariant under velocity
inversion $v \rightarrow -v$ for symmetric objects $F(\theta)=F(2\pi-\theta)$.
\begin{figure}[b]
\includegraphics[width=2.5in]{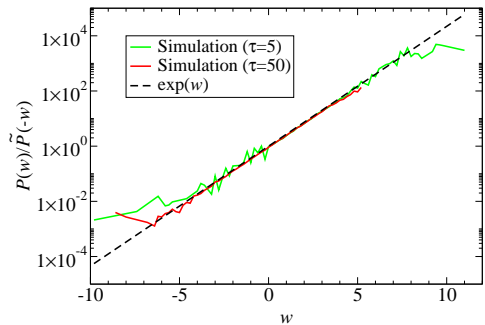}
\caption{Crooks relation (\ref{eq:crooks}) for a triangular object with 
velocity $v=0.2 (\tau=50)$ and $v=2 (\tau=5$).}
\label{fig:crooks}
\end{figure}

Having obtained the explicit form of the work distribution, we turn to the
verification of the fluctuation-dissipation relations. It will suffice to check
the Crooks relation which, in the present problem with $\Delta F =0$, reads
$P(w)=\exp(w)\tilde{P}(-w)$. The tilde here corresponds to velocity inversion. 
By Fourier transformation, the Crooks relation is equivalent
with the symmetry condition $\tilde{G}(-q-i)=G(q)$, 
which is indeed verified as, $
\tilde{G}(-q-i)-G(q)=-2\tau v\int_{0}^{2\pi}F(\theta)\sin \theta d\theta =0$.
The last equality follows from the fact that the object is closed.\newline The explicit result (\ref{gq}) allows, by expansion of $G(q)$, to evaluate the cumulants $\kappa_{n}$ of the random variable $w$:
\begin{widetext}
\begin{subequations}
\begin{align}
\kappa_{1}&=\langle w \rangle =\tau v \int_{0}^{2\pi} d\theta
F(\theta)\biggl\{
\frac{4v\sin^{2}\theta e^{-v^{2}\sin^{2}\theta}}{\sqrt{\pi}}+2\sin \theta
(1+2v^{2}\sin^{2} \theta)(1+\text{erf}[v\sin\theta ])
\biggr\}, \label{eq:k1} \\
\kappa_{2}&=\langle \delta w^2 \rangle=8\tau v^2 \int_{0}^{2\pi} d\theta
F(\theta)\sin^{2}\theta \biggl\{\frac{2e^{-v^{2}\sin^{2}
\theta}(1+v^{2}\sin^{2}\theta)}{\sqrt{\pi}}
+v\sin \theta (3+2v^{2}\sin^{2}
\theta )(1+\mbox{erf}[v\sin\theta]) \biggr\}, \label{eq:k2} \\
\kappa_{3}&=16\tau v^3 \int_{0}^{2\pi} d\theta F(\theta)\sin^{3}\theta
\biggl\{\frac{2e^{-v^{2}\sin^{2}
\theta}v\sin\theta(5+2v^{2}\sin^{2}\theta)}{\sqrt{\pi}}
+(3+12v^{2}\sin^{2} \theta +4v^{4}\sin^{4}
\theta )(1+\mbox{erf}[v\sin\theta]) \biggr\}, \label{eq:k3} \\
\kappa_{4}&=64\tau v^4 \int_{0}^{2\pi} d\theta
F(\theta)\sin^{4}\theta
\biggl\{\frac{2e^{-v^{2}\sin^{2}
\theta}(4+9v^{2}\sin^{2}\theta+2v^{4}\sin^{4}\theta)}{\sqrt{\pi}} \nonumber\\
&\hspace{1.6in}+v\sin\theta (15+20v^{2}\sin^{2} \theta +4v^{4}\sin^{4}
\theta )(1+\text{erf}[v\sin\theta]) \biggr\}. \label{eq:k4}
\end{align}
\end{subequations}
\end{widetext}
We next turn to two limits of particular interest.  In the quasi-static limit $v
\rightarrow 0$, expansion of $G(q)$ in $v$ leads to:
\begin{equation}\label{gqv}
G(q)\approx -\frac{8q(i+q)}{\sqrt{\pi}}v^{2}
\int_{0}^{2\pi} d\theta F(\theta)
\sin^{2}\theta + O(v^3).
\end{equation}
Hence $P(w)$ converges to $\delta(W)$ in the strict quasi-static limit
($G(q)=0$). Keeping the leading term in Eq. (\ref{gqv}), one concludes
that $P(w)$ is Gaussian. This level of perturbation corresponds to the linear
Gaussian regime around equilibrium. The average work reduces to the familiar Joule heating, which in original
variables, reads:  $\langle W \rangle \approx \gamma V^{2} t$. Here  $\gamma$
is the friction coefficient, the proportionality factor in friction force versus
speed: $F_{friction}=\gamma {V}$. The expression for the friction coefficient
$\gamma$ agrees with a direct calculation of this quantity
\cite{meurs04}, namely
$\gamma=4S\rho \sqrt{{k_{B}Tm}/{2\pi}}\int_{0}^{2\pi}F(\theta)\sin^{2}\theta
d\theta$. The second moment reproduces an equilibrium fluctuation-dissipation
relation. In original variables: $\beta \langle \delta W^2 \rangle=2 \langle W
\rangle$. The Jarzynski equality implies more generally for a Gaussian shape of
$W$ that  the so-called fluctuation-dissipation ratio $R=\beta \langle \delta W^2
\rangle/2(\langle W \rangle -W_{rev})$ be equal to one \cite{ritort03}.
In the present case, the reversible work, $W_{rev}=\Delta F$, is equal to zero. 
\newline The quasi-static limit has to be compared but also contrasted with  the long
time limit $\tau \rightarrow  \infty$.  We first note that all the
cumulants are proportional to $\tau$, a property characteristic for processes
with independent increments. As a result, one finds in the limit  $\tau
\rightarrow  \infty$ that $W$ converges to $\langle W \rangle$ (the law of large
numbers), and more precisely that the random variable $(W-\langle W
\rangle)/\sqrt{\langle \delta W^2 \rangle}$
converges to a normal random variable (central limit theorem).  We stress
however that, while in this case the dominant part of the probability mass is
indeed rendered correctly by a  Gaussian ansatz centered around  $\langle W
\rangle$, the validity of the fluctuation theorems rests on the contribution of  the so-called extreme non-Gaussian deviations, as all higher order cumulants contribute equally (all proportional to $\tau$) to the
Crooks relation. In particular, the average work is not related in any obvious
way to the free energy difference  nor does the fluctuation-dissipation ratio
verify the near equilibrium result $R=1$.
\setlength{\tabcolsep}{0.05in}
\squeezetable
\begin{table*}
\caption{Comparison between molecular dynamics simulation and theory
($v\tau=10$). Cumulants
$\kappa_1, \cdots,  \kappa_4$, fluctuation-dissipation ratio $R$,
and the Jarzynski average $\langle e^{-w} \rangle$ are shown for the disk (upper
part)
and triangle (lower part).} 
\label{tbl:moments}
\begin{tabular}{d|dd|dd|dd|dd|dd|d}
\toprule
\multicolumn{1}{c|}{\raisebox{-1.0ex}[0pt]{$v$}} &
\multicolumn{2}{c|}{$\kappa_1$}& \multicolumn{2}{c|}{$\kappa_2$} &
\multicolumn{2}{c|}{$\kappa_{3}$} & \multicolumn{2}{c|}{$\kappa_{4}$}  &
\multicolumn{2}{c|}{$R$} &
\multicolumn{1}{c}{$\langle e^{-w} \rangle$ }\\
&\multicolumn{1}{c}{Simulation}
&\multicolumn{1}{c|}{Theory}&\multicolumn{1}{c}{Simulation}
&\multicolumn{1}{c|}{Theory}&\multicolumn{1}{c}{Simulation}
&\multicolumn{1}{c|}{Theory}&\multicolumn{1}{c}{Simulation}
&\multicolumn{1}{c|}{Theory}&\multicolumn{1}{c}{Simulation}
&\multicolumn{1}{c|}{Theory}&\multicolumn{1}{c}{Simulation}\\
\colrule
0.01&0.190&0.226&0.394 &0.451&0.004&0.002&-0.009 &0.005&1.04 &0.998&   1.01\\
0.10&2.46&2.26&5.04 &4.62&0.051&0.042& 0.042 &0.053&1.03 &1.02&0.99\\
0.50&13.0&12.0&38.3&35.7&0.353&0.385& 0.191 &0.229&1.48&1.49&0.31\\
1.00&30.1&27.8&169.&160.&0.466&0.509&0.234&0.300&2.80&2.88&0.033\\
2.00&86.3&80.2&1310.&1280.&0.458&0.507&0.225&0.282&7.57&7.96&0.0043\\
\hline\hline
 0.01&0.078&0.078&0.156&0.153&-0.076&-0.111&0.012&0.018&1.004&0.981&1.00\\
-0.01&0.078& 0.079&0.159&0.160& 0.002& 0.111& 0.024&0.017& 1.01& 1.02& 1.00\\
\hline
 0.10&0.708&0.729&1.23&1.26&-0.356&-0.353& 0.181&0.193& 0.870& 0.865& 1.00\\
-0.10& 0.811& 0.843&1.86&1.96& 0.327& 0.342& 0.123&0.156& 1.15& 1.16& 1.00\\
\hline
 0.50& 2.73& 2.76&2.97&2.99&-0.566&-0.530& 1.04&1.03& 0.544& 0.542& 0.77\\
-0.50& 5.43&5.63&21.48&23.2&0.585&0.629& 0.371&0.464& 1.98& 2.06& 1.01\\
\hline
 1.00& 4.32& 4.14&5.05&4.14& 0.395& 0.189& 0.965&0.737& 0.583& 0.500& 0.53\\
-1.00&14.9&15.6&106.& 120.&0.623&0.679&0.383&0.511&3.57&3.85&1.00\\
\hline
 2.00& 6.09& 6.08& 10.9&11.0&0.614&0.627&0.417&0.446& 0.891& 0.903& 0.11\\
-2.00&48.5&52.0&855.&1040.&0.568&0.646&0.292&0.443&8.83&10.0&1.10\\
\botrule
\end{tabular}
\end{table*}
We finally turn to a comparison of the above analytic results with those from hard disk molecular dynamics for a dilute gas with $N = 2000$ disks of diameter $d= 1$ and mass $m = 1$ (see Refs. \cite{vandenbroeck04,vandenbroeck05} for the detailed simulation methods.) The initial positions and velocities of the disks are sampled from a microcanonical ensemble (initial  "temperature" $T = 1$) in a square box of L=1000 (i.e., initial gas density $\rho =0.002$) with periodic boundary conditions in both $x$ and $y$ directions.  Averages were taken over 400,000 realizations. Two simple shapes are considered for the object: a disk  with a diameter $A=10$ ($F(\theta)=1/2\pi$) and an isosceles triangle with a base length $A=10$ and an apex angle $\phi=20^\circ$ [$F(\theta)$ is the sum of three delta contributions at the angles $\theta=\pi-\phi$, $\theta=3\pi/2$ and $\theta=\phi$]. For the disk, we have by symmetry  that $P=\tilde P$. The latter distributions can however be very different for the triangle, see below. The triangle is placed in such a way that its symmetry axis is parallel to the $x$ axis, cf. Fig. \ref{fig:snapshots}. An illustrative comparison between analytical and simulation results is shown in Figs.~\ref{fig:pw}-\ref{fig:crooks} and Table~\ref{tbl:moments} for a wide range of values of  $v$ and  $\tau$. Overall, qualitative agreement is observed. In particular, the progressive change in the general shape of the probability distribution $P(w)$ from its quasi-static Gaussian shape to a very complicated multi-peaked distribution for the triangle is well reproduced by the present theory. The comparison of moments given in Table~\ref{tbl:moments} confirms this agreement. In the linear response regime ($R\sim 1$), the molecular dynamics simulation reproduce quite well the Jarzynski equality, meaning that such simulations could be used as an accurate estimate of the free energy difference (which happens to be zero here). However, as the velocity increases, the Jarzynski equality is violated because its validity rests on the contribution of extremely rare events.  Note however the following exception: the triangular object moving to the left satisfies the Jarzynski equality  surprisingly well even when its velocity is greater than the mean velocity of the gas particles.  The intuitive explanation is that negative work, corresponding to particles hitting the triangle while it moves away from them,  can be more easily realized by the collision on the elongated side of the triangle (see the left panel of Fig. \ref{fig:snapshots}). This observation is also supported at the level of the Crooks relation, cf.  Fig \ref{fig:crooks}, and is in agreement with a general argument (C. Jarzynski, private communication) that the Jarzynksi equality is verified more easily when operating at higher dissipation, cf. Table~\ref{tbl:moments}.

We conclude with a few relevant comments. Only in the quasi-static limit
(including the linear regime around equilibrium) is the convergence of $P(W)$ 
to a delta function and to a Gaussian strong enough to be allowed to interchange
this limit with the Jarzynski average. One concludes that $\langle W \rangle$
converges to $\Delta F$ and the fluctuation-dissipation ratio $R$ converges to
$1$. In the limit  $\tau \rightarrow \infty$, $W$, being the sum of independent
increments,  both law of large numbers and central limit theorem apply, but the
Jarzynski average is dominated by the extreme non-Gaussian fluctuations.   The
present study also illustrates that one has to go far out of equilibrium, in
case the speed of the object comparable to the speed of the particles,  to see
significant deviations from linear response theory.  Furthermore, one will, in
the latter case ($R$ significantly different from $1$),  typically observe 
violations of the Jarzynski equality because it becomes practically impossible
to sample extremely rare events.  We however found an interesting exception to
this rule:  for the motion of a highly asymmetric object, the negative work
contributions of the not-so-rare collisions on its elongated tail  allow to
verify the Jarzynski equality outside the regime of linear response.
Concomitantly, the Jarzynski equality can in such cases be used to estimate
free energy differences from far from equilibrium measurements. 

\bibliographystyle{apsrev}
\bibliography{joule}

\end{document}